\documentclass[sigconf]{acmart}

\AtBeginDocument{%
  \providecommand\BibTeX{{%
    \normalfont B\kern-0.5em{\scshape i\kern-0.25em b}\kern-0.8em\TeX}}}

\setcopyright{acmcopyright}
\copyrightyear{2022}
\acmYear{2022}
\acmDOI{10.1145/3470496.3527443}

\acmConference[ISCA '22]{The 49th Annual International Symposium on Computer Architecture}{June 18--22, 2022}{New York City, NY}
\acmPrice{15.00}
\acmISBN{978-1-4503-8610-4/22/06}

\usepackage{graphicx}
\usepackage{textcomp}
\usepackage{xcolor}
\usepackage{multirow}
\usepackage{indentfirst}

\usepackage[normalem]{ulem}
\newcommand{\tabincell}[2]{\begin{tabular}{@{}#1@{}}#2\end{tabular}} 
\usepackage{enumitem}
\usepackage{subfigure}
\usepackage{booktabs} 
\usepackage{caption}
\usepackage{stfloats}
\usepackage{algpseudocode}
\usepackage[flushleft]{threeparttable} %
\newcommand{\hr}[1]{\textcolor{black}{#1}}
\newcommand{\yang}[1]{\textcolor{black}{#1}}
\newcommand{\zy}[1]{\textcolor{black}{#1}}
\newcommand{\cheng}[1]{\textcolor{black}{#1}}

\newcommand{\name}{EyeCoD}

\begin{document}

\title{EyeCoD: Eye Tracking System Acceleration via FlatCam-based Algorithm \& Accelerator Co-Design}

\newcommand{\affilRICE}{$^1$}
\newcommand{\affilMeta}{$^2$}
\newcommand{\eq}{$^\ast$}

\author{
\texorpdfstring{
{Haoran You\affilRICE\eq}\quad
\authornote{These authors are co-first authors.}
{Cheng Wan\affilRICE\eq}\quad
{Yang Zhao\affilRICE\eq}\quad
{Zhongzhi Yu\affilRICE\eq}\quad
{Yonggan Fu\affilRICE}\quad
{Jiayi Yuan\affilRICE}\quad \\
{Shang Wu\affilRICE}\quad
{Shunyao Zhang\affilRICE}\quad
{Yongan Zhang\affilRICE}\quad
{Chaojian Li\affilRICE}\quad
{Vivek Boominathan\affilRICE}\quad \\
{Ashok Veeraraghavan\affilRICE}\quad
{Ziyun Li\affilMeta}\quad
{Yingyan Lin\affilRICE}\quad \\}
\affiliation{
    \begin{tabular}{cc}
        \affilRICE Rice University  & \affilMeta Meta Reality Labs
    \end{tabular}
}
    \begin{tabular}{c}
        \{hy34, chwan, zy34, zy42, yf22, jy101, sw99, sz74, yz87, cl114, vivekb, vashok, yingyan.lin\}@rice.edu  \\
        liziyun@fb.com
    \end{tabular}
\country{}
}

\def \authors{Haoran You, Cheng Wan, Yang Zhao, Zhongzhi Yu, Yonggan Fu, Jiayi Yuan, Shang Wu, Shunyao Zhang, Yongan Zhang, Chaojian Li, Vivek Boominathan, Ashok Veeraraghavan, Ziyun Li, Yingyan Lin}



\renewcommand{\shortauthors}{You et al.}

\begin{abstract}
     Eye tracking has become an essential human-machine interaction modality for providing immersive experience in numerous virtual and augmented reality (VR/AR) applications desiring high throughput (e.g., 240 FPS), small-form, and enhanced visual privacy. 
    However, existing eye tracking systems are still limited by their: (1) large form-factor largely due to the adopted bulky lens-based cameras; (2) high communication cost required between the camera and backend processor; and (3) potentially concerned low visual privacy, thus prohibiting their more extensive applications. To this end, we propose, develop, and validate a lensless FlatCam-based eye tracking algorithm and accelerator co-design framework dubbed \textbf{EyeCoD} to enable eye tracking systems with a much reduced form-factor and boosted system efficiency without sacrificing the tracking accuracy, paving the way for next-generation eye tracking solutions.
    \textbf{On the system level}, we advocate the use of lensless FlatCams instead of lens-based cameras to facilitate the small form-factor need in mobile eye tracking systems, which also leaves rooms for a dedicated sensing-processor co-design to reduce the required camera-processor communication latency.
    \textbf{On the algorithm level}, EyeCoD integrates a predict-then-focus pipeline that first predicts the region-of-interest (ROI) via segmentation and then only focuses on the ROI parts to estimate gaze directions, greatly reducing redundant computations and data movements.
    \textbf{On the hardware level}, 
    we further develop a dedicated accelerator that (1) integrates a novel workload orchestration between the aforementioned segmentation and gaze estimation models, (2) leverages intra-channel reuse opportunities for depth-wise layers, (3) utilizes input feature-wise partition to save activation memory size, and (4) develops a sequential-write-parallel-read input buffer to alleviate the bandwidth requirement for the activation global buffer. 
    On-silicon measurement and extensive experiments validate that our EyeCoD consistently reduces both the communication and computation costs, leading to an overall system speedup of 10.95$\times$, 3.21$\times$, and 12.85$\times$ over general computing platforms including CPUs and GPUs, and a prior-art eye tracking processor called CIS-GEP, respectively, while maintaining the tracking accuracy.
    Codes are available at \url{https://github.com/RICE-EIC/EyeCoD}.
\end{abstract}




\maketitle

\section{Introduction}







Eye tracking has emerged as a increasingly crucial component for various applications that require human-machine interactions, e.g., virtual and augmented reality (VR/AR) devices \cite{xie2021q_vr,zhao2021holoar,liu2019intelligent,ILLIXR}.
For example, Foveated Rendering (FR) \cite{kaplanyan2019deepfovea} is one of the core technologies that enables immersive user experiences in VR/AR applications requiring high-performance eye tracking. 
In particular, FR renders a high-resolution picture only in locations where users are looking at and a low-resolution one for the remaining background.
Despite their promise, existing eye tracking systems such as \cite{ASIC_gaze_processor} are still limited in their achievable throughput (e.g., still $<$ 30 FPS) and thus cannot fully satisfy the desired real time performance requirements, e.g., $>$ 240 FPS for supporting frequent and substantial human-machine interactions in mobile AR/VR devices of limited computing resources \cite{liu2019intelligent,michealiedm2021}.
The bottlenecks are three-fold:
\uline{First}, on the system level, previous eye tracking systems rely on lens-based cameras that have a large form-factor especially thickness and thus can only be placed far away from the backend processor, resulting in a high communication cost between the camera and processor and thus limiting the overall system latency;
\uline{Second}, on the data level, the captured images often contain a significant amount of redundancy as only a small portion of the images contains human eyes;
\uline{Third}, on the model level, current state-of-the-art award-winning solutions for both eye segmentation (e.g., OpenEDS2019 \cite{garbin2019openeds}) and gaze estimation (e.g., OpenEDS2020 \cite{palmero2021openeds2020}) require deep neural networks (DNNs) with paramount (e.g., up to 16G) FLOPs.

The above bottleneck analysis of existing eye tracking solutions has uniquely motivated our system design.
Specifically, for alleviating the aforementioned system-level inefficiency, lensless cameras \cite{FlatCam,merge_flatcam_chip_together,khan2020flatnet} have emerged as promising solutions. For example, FlatCam \cite{FlatCam} can be 5$\times$ $\sim$ 10$\times$ thinner and lighter than lens-based cameras by replacing the focal lenses with a coded binary mask, which encodes the incoming light instead of directly focusing it.
The encoded information of FlatCam's sensing measurements can be computationally decoded to reconstruct the captured images with potentially introduced artifacts and noises during the mask fabrication and measurement processes.
Furthermore, the reduced form-factor especially thickness leaves room for attaching the backend eye tracking processor to be closer to the front-end cameras, largely reducing the distance between the camera and processor and thus corresponding communication costs for reducing the overall system latency of eye tracking.
For tackling the data-level inefficiency, identifying the core eye area in the captured images can potentially reduce both a large amount of computational costs in the required gaze estimation model and corresponding data storage/movement costs of the eye tracking processor.
For the model-level inefficiency, a thorough algorithm and corresponding hardware accelerator design space exploration is crucial for largely improving the hardware utilization of eye tracking acceleration.

Motivated by the aforementioned bottleneck analysis and new opportunities, we advocate lensless camera based eye tracking systems for (1) alleviating the bottlenecks in existing eye tracking systems and (2) leveraging the aforementioned opportunities to largely enhance the achievable throughput of eye tracking systems, and make the following contributions:   

\begin{itemize}
    \item We propose a lensless FlatCam-based eye tracking algorithm and accelerator co-design framework dubbed EyeCoD, which aims to leverage FlatCam's much reduced form-factor to design a real-time eye tracking system (i.e., $>$ 240 FPS) by harmonizing both algorithm- and accelerator-level innovations. 
    Specifically, EyeCoD (1) explores the possibility of replacing lens-based cameras with lensless cameras featuring a thinner and lighter form-factor, yet without degrading the tracking accuracy, and (2) further accelerates both eye tracking computations and data movements with a dedicated accelerator attached to the lensless camera to largely reduce the overall system latency.
    
    \item On the algorithm level, EyeCoD integrates (1) a sensing-processing interface that directly encodes the first layer of eye tracking models to FlatCam's mask, and (2) a predict-then-focus pipeline that first predicts the region-of-interest (ROI) based on eye semantic segmentation and then only focuses on the ROI parts to estimate the gaze directions, largely reducing the redundant computations and data movements. 
    
    \item On the hardware level, EyeCoD further develops a dedicated accelerator that can be directly attached to FlatCam for accelerating eye tracking computations and data movements, by (1) enhancing data locality via dedicated workload orchestration between the eye segmentation (predict) and gaze estimation (focus) models; (2) exploring the reuse opportunity for depth-wise layers; and (3) leveraging activation partition and memory access parallelism to save on-chip storage and off-chip bandwidth, respectively.
    
    \item On-silicon measurements and extensive experiments validate the effectiveness of our proposed EyeCoD framework. Specifically, EyeCoD leads to 10.95$\times$, 3.21$\times$, and 12.85$\times$ overall system speedups over general computing platforms including CPUs and GPUs, and the prior-art eye tracking processor called CIS-GEP~\cite{ASIC_gaze_processor}, respectively, while maintaining the tracking accuracy.
    
\end{itemize}
\section{Related Works}

\textbf{Eye Tracking Algorithms.}
Existing eye tracking algorithms include both model- and appearance-based methods.
The former~\cite{EyeTab,3dmm} builds a geometric model for eyes to predict the corresponding gaze, including both 2D and 3D models that use near infrared (NIR) illumination to create corneal reflections to estimate the gaze vector. The latter~\cite{tan2002appearance} directly maps the raw pixels to the gaze angles. Appearance-based methods in general have surpassed model-based ones for eye tracking, especially when being equipped with advanced deep learning methods. 
Different DNN structures have been proposed to enhance the performance of gaze estimation. For example,
~\cite{in_the_wild} proposed the first DNN model for gaze estimation and ~\cite{Pictorial} further proposed a hybrid network integrating both hourglass~\cite{hourglass} and DenseNet~\cite{DenseNet} to leverage auxiliary supervision based on the gaze-map;
~\cite{asymmetric} introduced ARE-Net, which consists of two smaller modules to first find directions from each eye individually and then estimate the reliability of each eye, respectively;
~\cite{Monocular} also defined two convolutional neural networks (CNNs) to predict head and gaze angles, respectively.
In parallel, different processing pipelines have been developed with diverse focuses on the input features. For example,
~\cite{in_the_wild} utilized minimal context by only using grayscale eye images and head poses as inputs;
~\cite{gazecapture} developed a multi-model CNN to extract information from two single eye images, including face image and face grid, for aiding the following gaze estimation; and
~\cite{RT-GENE} built an ensemble on top of the features extracted by two eye patches and head pose vectors, and achieved superior performance on several datasets~\cite{RT-GENE,Image_Normalized}.

\textbf{Lensless FlatCam.}
As traditional lens-based cameras inevitably require a certain focal length, which prohibits their applications to edge devices with stringent requirements on the form-factor, various lens-less imaging systems have been developed to alleviate the size or thickness bottleneck caused by the lens by capturing an image of a scene without physically focusing the incoming light. 
Generally, lensless imaging systems capture the scene either directly on the sensor or after being modulated by a mask element. In the latter cases, commonly adopted masks include phase masks \cite{Stork2013LenslessUC,PhlatCam}, diffusers \cite{Antipa}, amplitude masks \cite{FlatCam,Shimano}, compressive samplers \cite{6738433}, and spatial light modulators \cite{Chi,10.1117}. 
Since directly replacing lens with the aforementioned masks will lead to muddled sensor captures without any resemblance to the scene, either a recovery process is required to transform the captured information to recognizable images or some dedicated functions are adopted to achieve end-to-end system goals without reconstructing corresponding recognizable images. 
From the privacy perspective, the captured unrecognizable images can better maintain the visual privacy compared with lens-based cameras. 
In this work, we adopt a specific lensless camera named FlatCam, which favors general uses and generates phase masks with desired sharp point-spread-functions (PSFs). In particular, phase masks in a FlatCam modulates the phase of incident light according to the principles of wave optics, which allow most of the light to pass through with a high signal-to-noise ratio (SNR). Therefore, FlatCam systems are particularly desirable for low light scenarios and photon-limited imaging, which is very suitable for eye tracking applications on VR/AR devices where human eyes are underexposed. 
\textbf{Eye Tracking Accelerators and DNN Accelerators.}
Various eye tracking systems with high energy-/latency-efficiency have been proposed for empowering the next-generation VR/AR devices. They are either built on top of commercial devices or supported by customized accelerators. 
For the former case, ~\cite{brousseau2018smarteye} presented an accurate infrared eye tracking system on a smartphone equipped with an infrared camera and illumination. 
For the latter case, ~\cite{bong20160} developed a CMOS image sensor based gaze estimation processor to reduce power consumption and ~\cite{hong20152} proposed a low-power single-chip gaze estimation sensor equipped with a novel column-parallel pupil edge detection circuit for supporting their proposed pupil edge detection algorithm, which can achieve a 2.9$\times$ power consumption reduction. 
A recent work~\cite{moon2021sub} designed the first 3D model-based gaze estimator hardware that consumes less than 1mW power and achieves latency of 1ms per frame. 
In parallel, driven by the success of DNNs in the eye tracking field, there has been an increasing interest in accelerating DNN-based eye tracking systems with customized DNN accelerators \cite{du2015shidiannao,chen2017eyeriss,10.1109/ISCA45697.2020.00073,10.1109/ISCA45697.2020.00082}. In particular, DNN accelerators have achieved impressive progress and often adopt a carefully designed memory hierarchy and PE arrays to maximize data-reuse opportunities and to enhance parallel processing via dedicated micro-architectures and algorithm-to-hardware mapping methods (i.e., dataflows). 
\begin{figure*}[t]
    \centering
    \includegraphics[width=\linewidth]{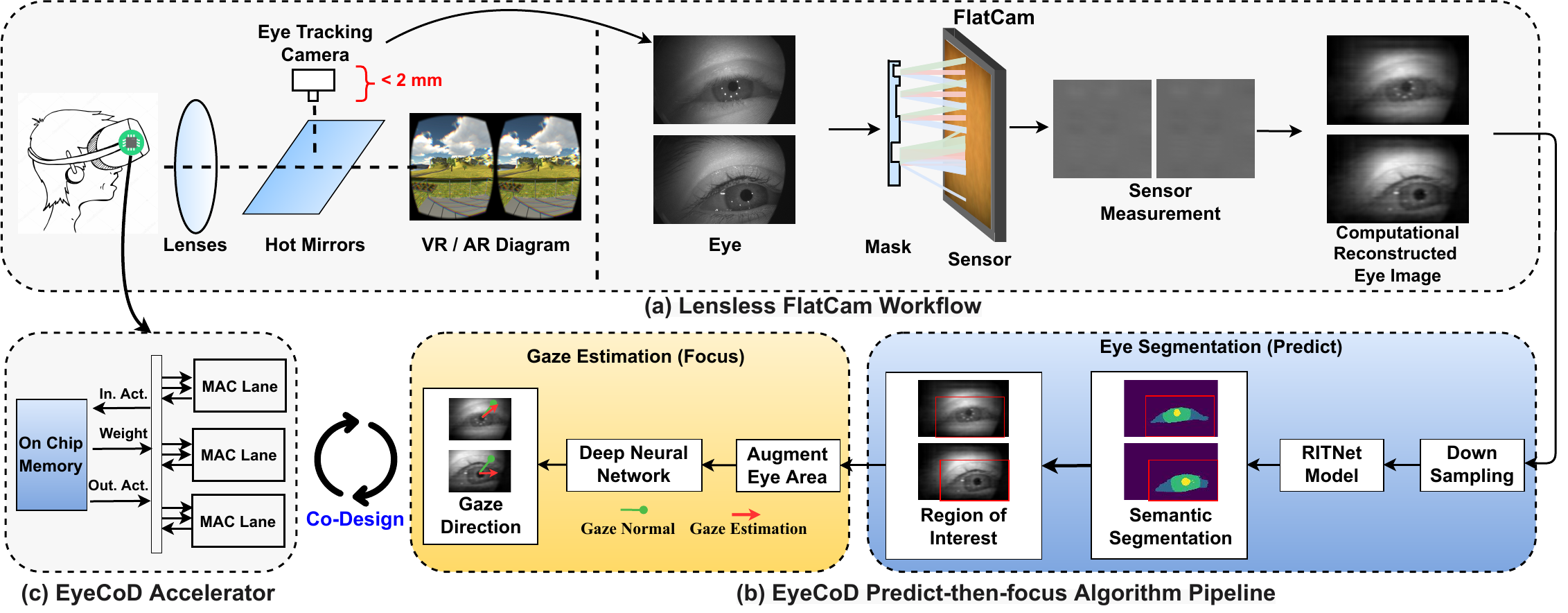}
    \caption{Overview of EyeCoD, an algorithm and accelerator co-design framework for end-to-end eye tracking acceleration.}
    \label{fig:overview}
\end{figure*}
For example, representative works, such as ShiDiannao \cite{du2015shidiannao} and Eyeriss \cite{chen2017eyeriss}, identified the performance bottleneck caused by the required massive data movements and proposed novel architectures and dataflows that aim to maximize data reuses for reducing the energy/time cost of accessing higher cost memories.

\section{EyeCoD: Motivation and Overview}

\subsection{Why Existing Eye Tracking Solutions Are Still Inefficient}

Eye tracking is known to be a core function for enabling high-quality immersive VR/AR experiences, and requires stringent requirements in terms of both real-time latency and high accuracy for gaze estimation \cite{palmero2021openeds2020}.
In general, there still exists a dilemma for designing eye tracking systems: On one hand, the end-to-end system latency needs to meet real-time performance, which desires compact end-to-end processing models/pipelines which can inevitably degrade the achieved tracking accuracy; 
On the other hand, adopting more complex processing models/pipelines favor the achievable tracking accuracy but can lead to a large system latency of performing eye tracking. For example, a state-of-the-art ASIC eye tracking processor \cite{ASIC_gaze_processor} implemented in an 65nm CMOS technology can only achieve a throughput of 30 FPS, limiting their more extensive applications \cite{michealiedm2021,liu2019intelligent,liu2020intelligent}.

To better understand the challenges associated with accelerating eye tracking systems, we analyze the bottlenecks from three levels of granularity:
(1) On the system level, current lens-based eye tracking camera requires a large form-factor, contradicting the desired small form-factor for mobile VR/AR applications with a head-mounted display (HMD), and thus the camera often locates far away from the central processor, resulting in a high communication cost between the camera and backend processor and thus limiting the achievable end-to-end latency \cite{DBLP:journals/corr/ChenJYSSVM16,michealiedm2021};
(2) On the data level, there remains a nontrivial amount of redundancy in the captured images, as only a small portion of which represents human eyes, and thus corresponding redundant acceleration costs.
(3) On the model level, current state-of-the-art award-winning solutions for both eye segmentation (e.g., OpenEDS2019 \cite{garbin2019openeds}) and gaze estimation (e.g., OpenEDS2020 \cite{palmero2021openeds2020}) require DNNs with paramount (up to 16G) FLOPs.
The above analysis regarding the inefficiency and bottleneck of existing eye tracking solutions has uniquely motivated our target dedicated algorithm and accelerator co-design framework for achieving both the real-time processing (e.g., $>$ 240FPS \cite{liu2019intelligent}) and the competitive tracking accuracy.


\subsection{Why EyeCoD Works and Overview}

Fig. \ref{fig:overview} shows an overview of the proposed EyeCoD framework, which integrates techniques from various system granularities dedicated to tackle the aforementioned three bottlenecks and thus can largely alleviate the dilemma between the achievable eye tracking efficiency and accuracy. 
On the system level, we advocate the use of lensless FlatCams instead of lens-based cameras to facilitate the small form-factor needed in mobile eye tracking systems, which also leaves rooms for a dedicated sensing-processor co-design to reduce the required camera-processor communication latency.
On the algorithm level, we leverage a predict-then-focus processing pipeline to first identify regions of interest (ROI) via periodic segmentation and then estimate the gaze direction only based on the extracted ROI, eliminating redundant data regions and corresponding algorithmic processing and data movements. 
Meanwhile, we explore eye tracking model design spaces and compression techniques on top of award-winning SOTA designs \cite{garbin2019openeds,palmero2021openeds2020} to additionally reduce algorithmic redundancy and corresponding acceleration cost.
Finally, we develop a dedicated accelerator to leverage the resulting properties from EyeCoD's system and algorithm level optimization, further improving the overall system efficiency.







\section{Proposed EyeCoD's Sensing and Processing Pipeline}

In this section, we present our EyeCoD's sensing and processing pipeline. Specifically, we first present (1) the preliminaries of lensless FlatCams in Sec. \ref{sec:flatcam}, (2) EyeCoD's sensing-processing interface in Sec. \ref{sec:interface}, and (3) EyeCoD's predict-then-focus processing pipeline and its model compression consideration in Sec. \ref{sec:predict_and_focus}.

\subsection{Preliminary of Lensless FlatCams}
\label{sec:flatcam}

\begin{figure}
    \centering
    \includegraphics[width=\linewidth]{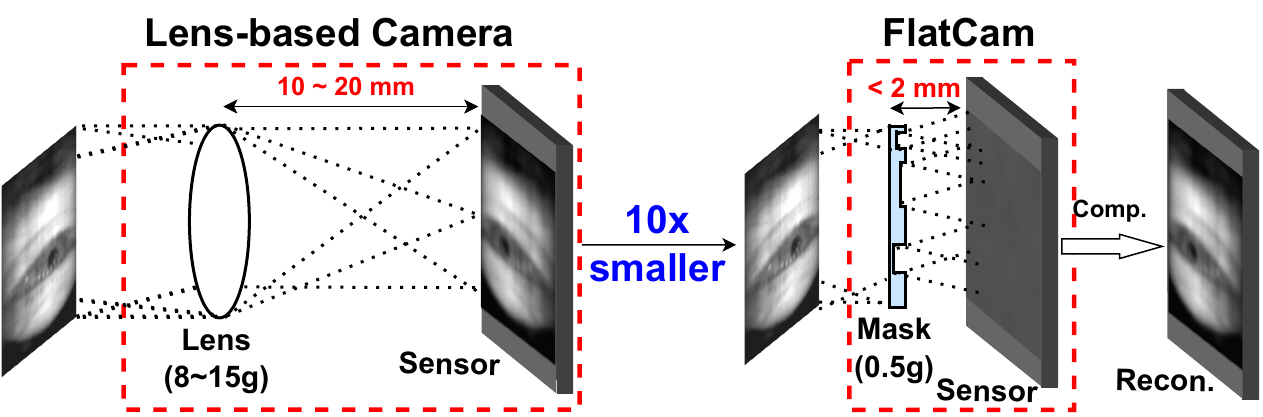}
      \caption{An illustrative comparison between lens-based cameras (left) and lensless FlatCam (right), where Comp. and Recon. denote computation and reconstruction, respectively.}
    \label{fig:flatcam}
\end{figure}

\textbf{Replacing Lens With a Lensless Coded Mask.}
Our EyeCoD advocates replacing the commonly adopted lens-based cameras in eye tracking systems with lensless cameras featuring both smaller thickness and weights (see Fig.~\ref{fig:flatcam}), in order to reduce (1) the camera-processor distance and (2) communication data volume (i.e., communicating intermediate features with smaller sizes instead of raw images with larger sizes) between the camera and processor, both leading to reduced communication cost between the camera and backend processor for eye tracking. In particular, while EyeCoD can potentially adopts various lensless cameras, in this work we consider 
FlatCam~\cite{asif2015flatcam} which replaces the bulky lens in lens-based cameras with a carefully designed thin mask placed on top of the conventional sensor array. During imaging, the incident light is first encoded by the mask and each pixel in the sensor measurement records a linear combination of light from multiple directions. The imaging process can be formulated as follows:

\begin{equation}
    y = \Phi_L x \Phi_R^T + e, 
\label{eq:imaging}
\end{equation}

\noindent where $x$ and $y$ are the input light and measurement, respectively, and $\Phi_L$ and $\Phi_R$ are transfer matrices representing the coded masks and $e$ captures the sensor noise. 
Replacing lens with thin masks of FlatCams can reduce the form-factor by orders of magnitude \cite{asif2015flatcam}, making it naturally suitable for mobile eye tracking systems (e.g., VR/AR devices), where there are stringent requirements on the devices' thickness and weight.

\textbf{Image Reconstruction.}
As shown in Fig.~\ref{fig:flatcam}, the sensing outputs of FlatCams do not capture the target scene but its computational combination that often does not convey readable information. 
To facilitate the following gaze estimation, we first reconstruct the scene image (i.e., captured eyes) by solving an inverse problem of the imaging process with a $\mathcal{L}_2$ norm regularization to reduce the noise during imaging, following ~\cite{asif2015flatcam}.
Specifically, the optimization goal of this image reconstruction can be formulated as:
\begin{equation}
    \arg \min_X \Vert\Phi_L X \Phi_R^T - y\Vert_2^2 + \epsilon \Vert X\Vert^2_2, 
\label{eq:flatcam_recon}
\end{equation}
where $\epsilon > 0$ is a regularization parameter and we optimize the reconstructed images $X$ by minimizing the above least-square objective function following \cite{FlatCam} towards obtaining the optimally reconstructed images $X_{rec}$. 

\textbf{Opportunities.} From the aforementioned background regarding FlatCam, we can see that as compared to eye tracking systems with lens-based cameras, lensless camera based ones exhibit a great potential in terms of smaller form-factor (e.g., 5$\times$ $\sim$ 10$\times$ thinner and $>$10$\times$ lighter), 
reduced communication costs between camera sensors and backend processors,
and improved visual privacy, leading to a reduced end-to-end system latency.

\subsection{EyeCoD's Sensing-processing Interface}
\label{sec:interface}

To leverage the aforementioned opportunities offered by lensless cameras, EyeCoD's sensing-processing interface 
replaces both FlatCam sensing and the first layer of the following eye tracking model with direct optical edge filtering using FlatCam's coded masks, similar to \cite{chen2016asp,Fu_2021_ICCV}, i.e., the coded masks' optical response emulates the first layer of the following DNNs. 
Such a sensing-processing interface offers two-fold benefits that are highly desirable for mobile eye tracking systems:
(1) the enabled sensing-processing co-design within the lensless cameras leads to FLOPs and acceleration cost savings, thanks to the first-layer optical computation, and thus requires a lower electronic power consumption \cite{chen2016asp}, especially for the UNet-like segmentation models \cite{chaudhary2019ritnet} of which the first layer has to process images of the highest resolution; and 
(2) embedding the first-layer of the following eye tracking model into FlatCam's coded masks favors reduced sensing-processor communication volume, since the intermediate sensor measurements now enjoy reduced sizes/channels as compared to the raw images captured by lens-based cameras.

\subsection{EyeCoD's Predict-then-focus Processing Pipeline}
\label{sec:predict_and_focus}

\begin{figure}
    \centering
    \includegraphics[width=0.95\linewidth]{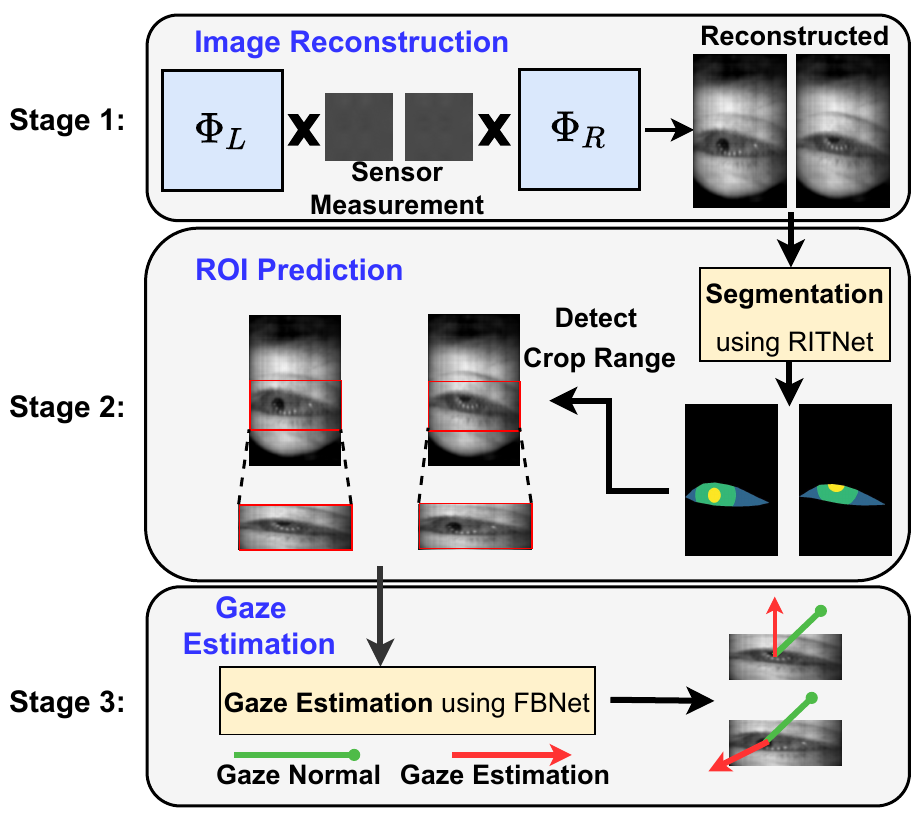}
    \caption{An overview of the proposed predict-then-focus processing pipeline.}
    \label{fig:alg_pipeline}
\end{figure}

\textbf{Pipeline Overview.}
EyeCoD's processing pipeline consists of three stages as shown in Fig~\ref{fig:alg_pipeline}: (1) image reconstruction as described in Sec. \ref{sec:flatcam}, (2) ROI prediction, which aims to predict the ROI centered around the human pupil in each reconstructed image, and (3) gaze estimation, which estimates the gaze based on the ROI derived from the previous stage.
Our new contribution lies in the intersection between the second and third stages, where we aim to precisely predict and crop the most informative core eye area (i.e., pupil, iris, and sclera) to estimate the gaze with 
lower costs. 
Note that the ROI prediction will only needed once for every 50 frames leveraging the fact that the movement of eyes are much slower than the movement of gaze directions \cite{palmero2021openeds2020}, while the gaze estimation will be continuously processed for each frame based on the latest predicted ROI.
Note that the costs of ROI prediction are amortized across 50 frames, therefore, the dependent gaze estimation is operating on an ROI extracted 50$\sim$100 frames ago.

\textbf{ROI Prediction. }
Not all pixels in each image are of the same importance to the corresponding gaze estimation.
Ideally, the ROI should contain a small area with pupil, iris and sclera in the center to provide sufficient information with the lowest resolution size for estimating the gaze. 
However, in the captured image, skin consumes a large portion of area, which have little information for gaze estimation but consumes considerable costs during inference, providing us with a promising source of redundancy.

To reduce computational overhead, we propose to predict the ROI and then only estimate gaze based on the extracted ROI accordingly. 
We first use a segmentation model to segment the core eye area, of which the advantage is that it favors diverse downstream tasks including gaze estimation and thus general uses.
However, the high noise in FlatCam reconstructed images (especially the sclera part) makes it more challenging for the segmentation model to precisely predict the whole core eye area than lens-based camera captured images. To address this issue,
directly using segmented core eye areas as the feature for ROI prediction is likely to lead to an inaccurate result, degrading the model accuracy of gaze estimation.

Luckily, we observe that pupils have a significantly different feature than the other parts in the image, as the pupil is usually a circle with a darker color than its surrounding. Therefore, the segmentation model can correctly segment the pupil with a high confidence. Furthermore, as pupils are normally located near the center of human eyes, we propose to use the segmented pupil center as an anchor for generating the ROI. Specifically, we predict the ROI by cropping a rectangle patch centered around the pupils, where the rectangle patch's width and height are of 1.5$\times$ more than the average width and height of the segmented sclera area 
to cover core eye areas according to the statistics of the adopted training dataset.
The predicted ROI is then passed to the gaze estimation model for generating the final eye tracking output, i.e., gaze vectors represented in a 3D coordinate system.

\section{Proposed {\name}'s Accelerator}\label{sec:accelerator}
This section introduces our {\name}'s accelerator design. In Sec.~\ref{sec:accelerator_challenges_oppotrunities}, we first analyze the challenges brought about by EyeCoD's predict-then-focus processing pipeline to derive the design principles of the accelerator design for further minimizing the processing latency and maximizing energy efficiency, and then describe the proposed accelerator with dedicated optimizations in Sec. \ref{sec:accelerator_architecture}. 

\subsection{Design Challenges and Principles }\label{sec:accelerator_challenges_oppotrunities}


\textbf{Design Challenges.}
As mentioned in Sec.~\ref{sec:predict_and_focus}, EyeCoD's predict-then-focus processing pipeline consists of two DNN models, 
\hr{i.e., a segmentation model for ROI prediction and a gaze estimation model for predicting gazes (see Fig.~\ref{fig:alg_pipeline}),}
to collaboratively construct an end-to-end eye tracking pipeline, aiming for both higher eye tracking accuracy and model efficiency. 
This means \hr{that} EyeCoD's accelerator is required to efficiently accelerate both of the above segmentation and gaze estimation models that feature diverse model structures as well as layer types and shapes. 
Hence, it brings about four challenges for effective hardware acceleration to deliver the desired eye tracking latency and efficiency towards practical 
\hr{deployment on} mobile devices with both constrained computation and memory resources. 


\textbf{Challenge \# I: Workload Orchestration between Segmentation and Gaze Estimation.}
As introduced in Sec.~\ref{sec:predict_and_focus}, EyeCoD's predict-then-focus processing pipeline processes the segmentation and gaze estimation models in parallel for reducing the overall latency of eye tracking, where the gaze estimation model continuously operates on every frame while the eye segmentation model executes once out of every $N$ frames ($N=50$ in our design to balance the \hr{achieved} eye tracking accuracy and imposed latency and energy costs (see Sec.~\ref{sec:alg_exps})). As such, proper workload orchestration should be considered \hr{in our} hardware acceleration design.

\begin{figure}
    \centering
    \includegraphics[width=\linewidth]{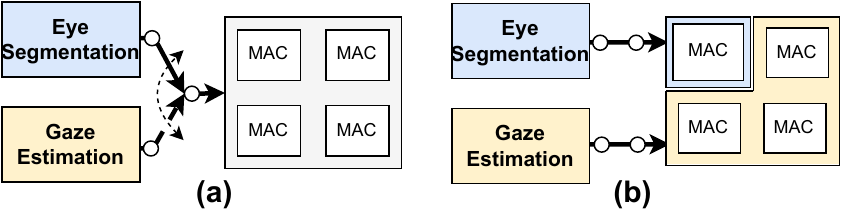}
     \caption{Two classical workload orchestration modes: (a) the time-multiplexing mode and (b) the concurrent mode, for accelerating both the eye segmentation and gaze estimation models.}
    \label{fig:workload_orchestration}
\end{figure}

Potentially, two classical workload orchestration modes, i.e., time-multiplexing and concurrent \hr{modes}, as shown in Fig.~\ref{fig:workload_orchestration}, can be adopted to orchestrate the eye segmentation and gaze estimation models on the same accelerator. 
However, these two workload orchestration modes either require a larger amount of computation resources (i.e., the time-multiplexing mode) or less opportunities for data reuses (i.e., the concurrent mode). 

(1) Timing-multiplexing mode. When utilizing the time-multiple-xing mode in Fig.~\ref{fig:workload_orchestration} (a), only one of the two models' layers occupies the accelerator's computation resources at a given time. As such, to ensure the performance of accelerating the bottleneck layers (i.e., layers with the largest number of operations (FLOPs)) which dominates the overall processing latency, adopting a time-multiplexing mode demands a larger amount of acceleration resources than the models' theoretical requirement. 
For better understanding, we provide an analysis here. As our eye segmentation's RITNet~\cite{chaudhary2019ritnet} and gaze estimation's FBNet-C\zy{100}~\cite{wu2019fbnet} contain \yang{140M and 1.06G FLOP}s, respectively, the theoretical computation resource requirement is 1024 multiplication-and-accumulations (MACs), if assuming the target eye tracking throughput is 240 FPS at a processing frequency of 350MHz. 
However, 256 additional MACs (i.e., corresponding to 25\% extra MACs if considering a theoretical requirement of 1024 MACs) are required to maintain the target 240 FPS system latency, when running the bottleneck layers \yang{(i.e., the third, fifth, forty-second, and forty-forth layer)} of the eye segmentation model~\cite{chaudhary2019ritnet} \yang{although the eye segmentation model executes once out of every 50 frames}.


(2) Concurrent mode. Considering a concurrent mode for EyeCoD's predict-then-focus processing pipeline, an accelerator spatially executes both the eye segmentation and gaze estimation models simultaneously in two fixed partitions of the accelerator's MACs as shown in Fig.~\ref{fig:workload_orchestration} (b) for a given cycle. \yang{Different from the time-multiplexing mode, the execution latency of bottleneck layers of the segmentation model are amortized to every 50 frames and do not dominate the overall processing latency. 
However, the concurrent mode brings about the drawback of less data reuse opportunities, since each of the two partitions has a \hr{reduced} amount of computation resources as compared to using all resources without partition (i.e., the time-multiplexing mode).} Naturally, a good partition scheme should balance the two models' complexity and execution frequency, which in our case will lead to only 4 MACs 
out of 1024 MACs being assigned to {\name}'s eye segmentation model and thus result in extremely less reuse opportunities and poor efficiency.


\textbf{Challenge \# II: Support for Various Layer Types}.
EyeCoD's predict-then-focus processing pipeline \hr{consists} of RITNet~\cite{chaudhary2019ritnet} for eye segmentation and FBNet-C\zy{100}~\cite{wu2019fbnet} for gaze estimation, \hr{that include} generic/point-wise/depth-wise convolution layers, fully-connected (FC) layers, and matrix-matrix-multiplication layers. 
The efficient processing of various layer types is a design challenge that needs to be addressed. 
In the following discussion, we compare the number of computation operations of different layer types to find the dominant layer types; analyze the reuse opportunity among various layer types; and show that depth-wise convolution layers require specific optimization. 

\begin{figure}
    \centering
    \includegraphics[width=0.85\linewidth]{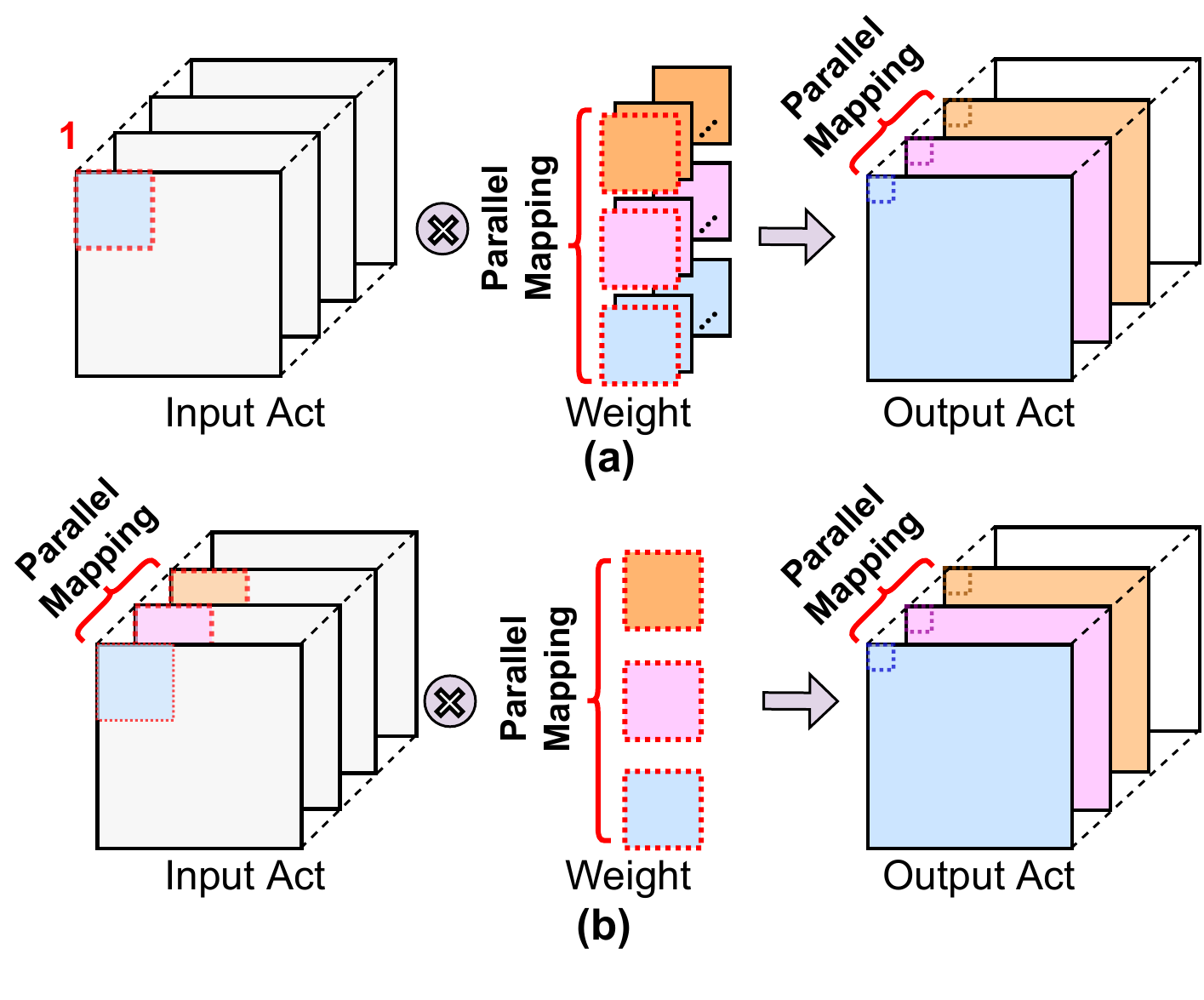}
    \caption{A computation illustration of (a) generic/point-wise convolution and (b) depth-wise convolution layers.}
    \label{fig:2_convolution}
\end{figure}

(1) Dominate layer type analysis. Considering a 50 frame processing \hr{of} EyeCoD's predict-then-focus processing pipeline when the eye segmentation is \hr{needed once}, 
generic convolution, point-wise convolution, depth-wise convolution, FC, and matrix-matrix multiplication layers account for 8.8\%, 68.8\%, 7.9\%, 0.001\%, and 14.5\% of the overall number of computation operations, respectively. 
FC is not the dominant layer, since it only accounts for about 0.001\% of the overall operations. 
Matrix-matrix multiplication, can be treated as point-wise convolution layer with 
\hr{a large batch size (i.e., $>1$). }
Therefore, generic convolution, point-wise convolution, and depth-wise convolution are the three dominant layer types in EyeCoD's predict-then-focus processing pipeline. 

(2) Reuse opportunity analysis. Fig.~\ref{fig:2_convolution} (a) and (b) illustrate the computation of generic/point-wise convolution and depth-wise convolution layers. 
There are two kinds of reuse opportunities shared by all convolution layers: \textit{Psum reuse} where the partial sums (Psums) are accumulated to calculate the corresponding output activation (Act) and
\textit{intra-channel reuse} where one input channel of weights \yang{are reused} by the corresponding channel of input activations to get the Psums (generic/point-wise convolution layers) or output activations (depth-wise convolution layers). Compared with depth-wise layer, generic/point-wise layer has input reuse where one input activation is reused by all 3D weight filters.

(3) Specific optimization requirement for depth-wise convolution layer.
Due to the limited reuse opportunities in depth-wise convolution layer, utilizing the same design for both generic/point-wise and depth-wise convolution layers typically leads to a very low \hr{MAC} utilization or requires a much higher input activation memory bandwidth (see Fig.~\ref{fig:2_convolution}). 
\hr{Our analytical analysis shows that all depth-wise convolution layers account for only 7.9\% of the overall number of computation operations, but consume 33.6\% of the overall processing time if using the same design as the generic/point-wise layers.}
Therefore, specific optimizations are necessary for \hr{the} depth-wise layer \hr{to fulfil the goal of real-time performance.}  

\textbf{Challenge \# III: Workload Partition to Save Activation Memory Size.}
If using the vanilla layer-by-layer processing, the theoretical on-chip activation memory size should fit the maximum requirement of each layer, \hr{i.e., \yang{2.78MB}, where} the eye segmentation model and the gaze estimation model occupy \yang{2.08MB and 0.70MB}, respectively. 
The \yang{2.78MB} on-chip memory size is unacceptable for the eye tracking application; let alone we only count the activation memory. 
Therefore, proper workload partition is required so that we only need to allocate the activation memory size for the processing of each individual partition.

\begin{figure}[t]
    \centering
    \includegraphics[width=1\linewidth]{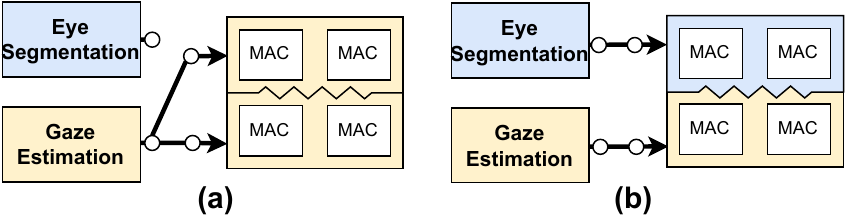}
    \caption{The proposed partial time-multiplexing mode for the workload orchestration of the eye segmentation and gaze estimation models: (a) the gaze estimation model occupies the computation resources; and (b) the eye segmentation and gaze estimation models run simultaneously.}
    \label{fig:proposed_workload_orchestration}
\end{figure}

\textbf{Challenge \# IV: High Activation Memory Bandwidth Requirement.}
As discussed in Challenge \# II, depth-wise convolution layers require a higher activation bandwidth to achieve a satisfying \hr{MAC} utilization (i.e., 32$\times\sim$128$\times$ higher bandwidth than the processing of processing generic/point-wise convolution layers for >50\% MAC utilization in our design). However, simply enlarging activation bandwidth leads to the bandwidth waste when processing other layers and increases the memory accesses cost. {\name}'s accelerator design should \hr{be optimized} to 
\hr{alleviate the stringent requirement of}
activation memory bandwidth for a better trade-off between depth-wise layers' and other layers' workloads.

\textbf{Design Principles.} Based on the above challenge analysis, we propose the following principles to take full advantage of EyeCoD's predict-then-focus processing pipeline for developing and optimizing the dedicated accelerator. In Sec.~\ref{sec:accelerator_architecture}, we 
\hr{incorporate} these design principles in our {\name} accelerator design.

\begin{figure}[t]
    \centering
    \includegraphics[width=1\linewidth]{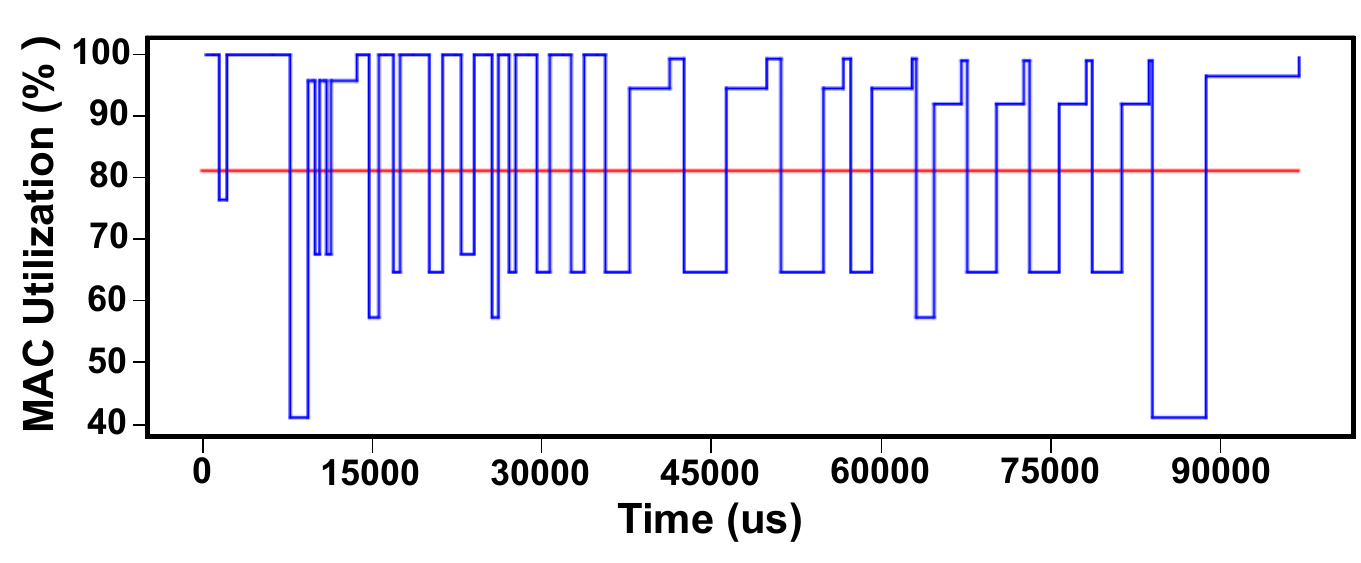}
    \caption{An illustration of the MAC utilization when running the gaze estimation model.}
    \label{fig:computation_utilization_FBNet}
\end{figure}

\textbf{Principle \# I: Partial Time-multiplexing Mode for Workload Orchestration.}
As the time-multiplexing and concurrent modes are not optimized for EyeCoD's predict-then-focus processing pipeline. 
\hr{Adopting them will} result in either a larger amount of computation resources or less reuse opportunities, we take advantage of both two modes and propose a partial time-multiplexing mode as shown in Fig.~\ref{fig:proposed_workload_orchestration}. 
In the proposed partial time-multiplexing mode, the gaze estimation model can fully occupy the computation resources as \hr{shown} in Fig.~\ref{fig:proposed_workload_orchestration} (a); or we can run the eye segmentation model and the gaze estimation model simultaneously as \hr{shown} in Fig.~\ref{fig:proposed_workload_orchestration} (b). Thanks to the simultaneous processing of both eye segmentation and estimation models (see Fig.~\ref{fig:proposed_workload_orchestration} (b)), the processing latency of the segmentation model's bottleneck layers are amortized to every 50 frames as in the concurrent mode, which tackles the large computation resource drawback in the time-multiplexing mode. 
\yang{Our evaluation shows that the proposed partial time-multiplexing mode has a 2.31$\times$ peak speedup than the time-multiplexing mode with a 10\% higher activation global buffer (GB) bandwidth and no computation resource (i.e., MAC) overhead.}
In addition, the partial time-multiplexing mode provides us the opportunity to better balance the reuse opportunities, the two models' complexity, and the two models' execution frequency to tackle the less reuse opportunity drawback in the concurrent mode. 
In particular, when the gaze estimation model requires a large amount of computation resources (i.e., generic/point-wise convolution layers), it fully owns the computation resources as shown in Fig.~\ref{fig:proposed_workload_orchestration} (a). On the other hand, only when the gaze estimation consumes a smaller amount of computation resources (i.e., depth-wise layers), we assign the unused resources to the eye segmentation model to run them simultaneously as shown in Fig.~\ref{fig:proposed_workload_orchestration} (b). At the same time, the eye segmentation model owns a larger amount of resources than that in the concurrent mode to enable a high reusability.

\textbf{Principle \# II: Intra-channel Reuse for Depth-wise Layer.}
As the depth-wise layer optimization is critical to speed up the gaze estimation model, the usually-unexplored intra-channel input reuse in generic/point-wise convolution layers need to be explored for depth-wise layer \yang{for achieving a high MAC utilization with an acceptable activation memory bandwidth and thus reducing the overall processing time}. 
Our evaluation shows that the proposed intra-channel reuse optimizations can reduce the processing time of depth-wise layers by 71\%. The detail description of the proposed intra-channel reuse optimizations is \hr{elaborated} in Sec.~\ref{sec:accelerator_architecture}. 
It should be note that the intra-channel reuses are limited for the layers with a stride of 2 and the last several layers with smaller input activation feature maps (e.g., 7/7 for the height/width of the input feature maps) in the gaze estimation model. As such, further increasing the MAC utilization of these layers is challenging. Thanks to the proposed partial time-multiplexing mode for workload orchestration, we tackle the MAC utilization challenge of these layers by running the segmentation model on the unused MACs. 
Fig.~\ref{fig:computation_utilization_FBNet} shows the MAC utilization when running gaze estimation model alone on {\name}'s accelerator. When the utilization is less than 80\% (i.e., the red line in Fig.~\ref{fig:computation_utilization_FBNet}), we can run the eye segmentation model on the unused resources in the proposed partial time-multiplexing mode for a >90\% overall MAC utilization. 

\begin{figure}[t]
    \centering
    \includegraphics[width=0.85\linewidth]{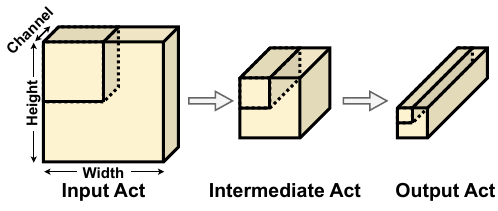}
    \caption{An illustration of input feature-wise partition for cross-layer processing.}
    \label{fig:cross_layer}
\end{figure}

\textbf{Principle \# III: Input Feature-wise Partition to Save Activation Memory Size.}
To save the activation memory size, we can partition the input image along the input activation's feature map dimensions (i.e., the height and the width of the input feature map) and process each individual partition through cross-layer processing. 
\hr{As illustrated in Fig.~\ref{fig:cross_layer},}
the on-chip activation memory only needs to store the activations of each partition. 
The overall activation memory size is about \yang{36\% (i.e., 1MB)} of that before partition.

\textbf{Principle \# IV: Parallelism of Memory Access and Processing to Save Activation Memory Bandwidth.}
Note that loading the activations from the memory for the next round of processing and the running of the current round of processing can be paralleled, where we denote the one round of processing as the processing using the same activations. Since each round of processing usually takes \yang{several cycles 
(e.g., the number of cycle equals to kernel sizes in our design)}, we propose to load the next-round activations sequentially from the memory during the current-round processing and then the already loaded next-round activations can be read out in parallel for the next round of processing. This parallelism of memory access and processing can save activation memory bandwidth. A sequential-write-parallel-read input activation buffer is needed to enable the parallelism, which is described in Sec.~\ref{sec:accelerator_architecture}.
Assuming a commonly-used 3$\times$3 kernel, the propose parallelism saves 50\%$\sim$60\% memory bandwidth and the sequential-write-parallel-read input activation buffer incurs a negligible area overhead of 0.58\%.


\subsection{Architecture of EyeCoD's Accelerator}\label{sec:accelerator_architecture}
This section describes the proposed accelerator architecture as well as the design optimizations following the principles in Sec.~\ref{sec:accelerator_challenges_oppotrunities}.

\begin{figure}[t]
    \centering
    \includegraphics[width=1.0\linewidth]{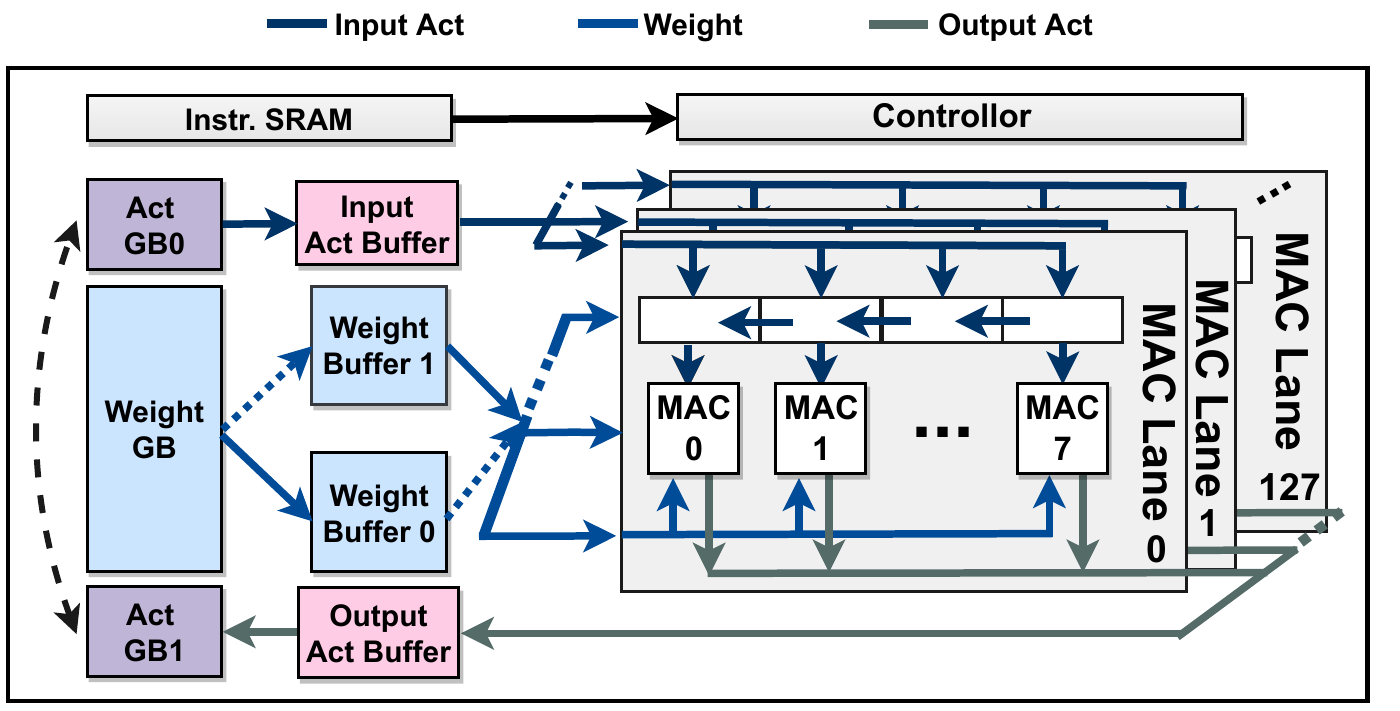}
    \caption{An illustration of {\name}'s accelerator.}
    \label{fig:arch}
\end{figure}

\textbf{Architecture Overview. }
Fig.~\ref{fig:arch} presents the architecture of EyeCoD's accelerator which consists of the following components: (1) on-chip memories for weights, input/output activations, and instructions, (2) computation resources, i.e., 128 MAC lanes, and (3) an on-chip controller. 
\uline{First}, two memory hierarchies are adopted. Specifically, the weight GB (i.e., global buffer) stores the parameters of the involved models as well as the reconstruction in EyeCoD's predict-then-focus processing pipeline. Two weight buffers are inserted between weight GB and MAC lanes and work in a "ping-pong" manner to avoid the weight load stalls. Similarly, an input Act buffer and a output Act buffer are inserted between the Act GBs and the MAC lanes to prepare activations for the MAC lanes or Act GB to eliminate input load or output write stalls. 
\uline{Second}, each MAC lane is composed of eight MACs and one input Act FIFO to store one row of input activations. The weights of one row are fetched one-by-one from the weight buffer for multiplying one row of input activations in the input Act FIFO. Therefore, each MAC lane is able to \hr{reuse the loaded input activation row, i.e., row-wise intra-channel reuse} which is ubiquitous among all convolution layers.
\uline{Third}, to implement EyeCoD's predict-then-focus processing pipeline, the on-chip controller reads instructions from the instruction SRAM to control the accelerator. 

\begin{figure}[b]
    \centering
    \includegraphics[width=0.95\linewidth]{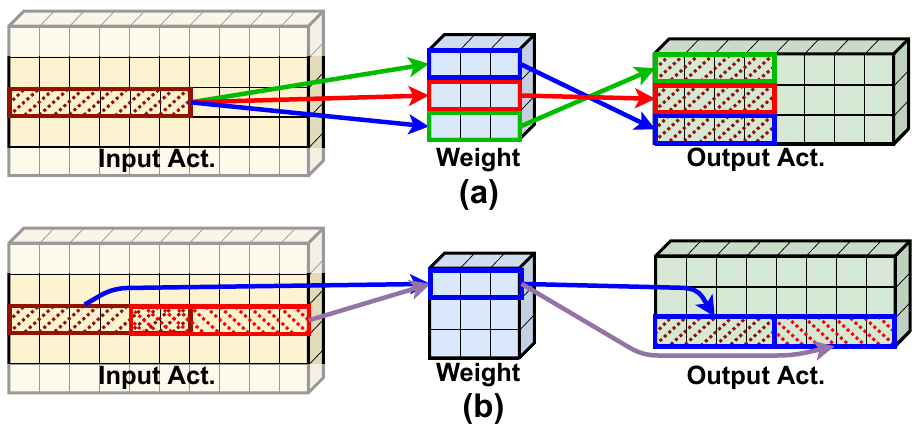}
    \caption{Optimizations for depth-wise convolution layers: (a) column-wise intra-channel reuse and (b) deeper row-wise intra-channel reuse.}
    \label{fig:optimization_DW}
\end{figure}

\textbf{Optimizations for Depth-wise Layer.}
As discussed in Sec.~\ref{sec:accelerator_challenges_oppotrunities}, the dataflow for generic/point-wise convolution layers is not sufficient for depth-wise layers, resulting in low MAC utilization or higher input Act bandwidth. To address this, intra-channel reuse is adopted for increasing MAC utilization and better leveraging the limited memory bandwidth. 
Fig.~\ref{fig:optimization_DW} illustrates two intra-channel reuse opportunities, \textit{column-wise intra-channel reuse} (Fig.~\ref{fig:optimization_DW} (a)) and \textit{deeper row-wise intra-channel reuse} (Fig.~\ref{fig:optimization_DW} (b)), for depth-wise convolution layers besides the row-wise intra-channel reuse on each MAC lane. 
For the former \textit{column-wise intra-channel reuse} method, multiple weight rows in one column of each 3D weight filter reuse one same input activation row and generate multiple output activation rows in the corresponding column. This technique achieves an utilization improvement proportional to the number of available weight rows or the kernel size (e.g., 3 or 5 for the gaze estimation model) as shown in Fig.~\ref{fig:optimization_DW} (a).
The latter \textit{deeper row-wise intra-channel reuse} is proposed because the row-wise intra-channel reuse on each MAC lane is limited by the number of MACs of each MAC lane (i.e., 8 MACs/MAC lane in our design). For the latter \textit{deeper row-wise intra-channel reuse}, we tile one input Act row into two sub rows, 
and further spatially map these two sub input Act rows and their corresponding weight row to two MAC lanes, doubling the MAC utilization.
\begin{figure*}
    \centering
    \includegraphics[width=0.95\linewidth]{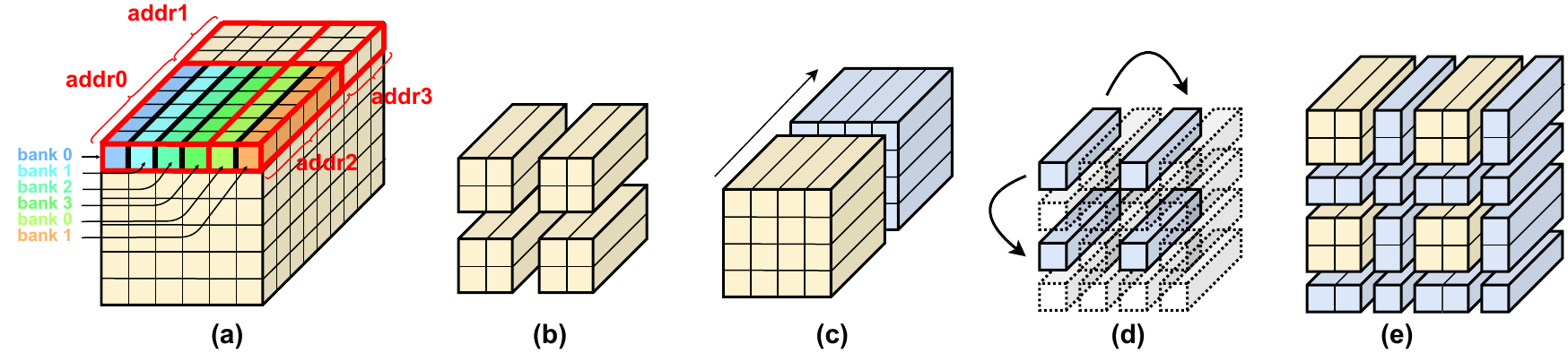}
    \caption{An illustration of the proposed activation GB storage arrangement: an example of (a) the storage arrangement of a 6$\times$6$\times$24 activation tensor, (b) the partition operation, (c) the concatenation operation, (d) the downsampling operation, and (e) the upsampling operation. }
    \label{fig:act_GB}
\end{figure*}

\textbf{Activation GB Storage Arrangement.}
Due the diverse model structures as well as layer types and shapes in EyeCoD's predict-then-focus processing pipeline, various activation reshaping operations are needed.
The various reshaping operations impose a challenge for the activation GB storage arrangement, i.e., how to support different activation reshaping operations without complicating controls.
We first classify the activation reshaping operations into four classes and then propose an optimized activation GB storage arrangement considering the characterizations of the reshaping operations.

\begin{figure}[t]
    \centering
    \includegraphics[width=1.0\linewidth]{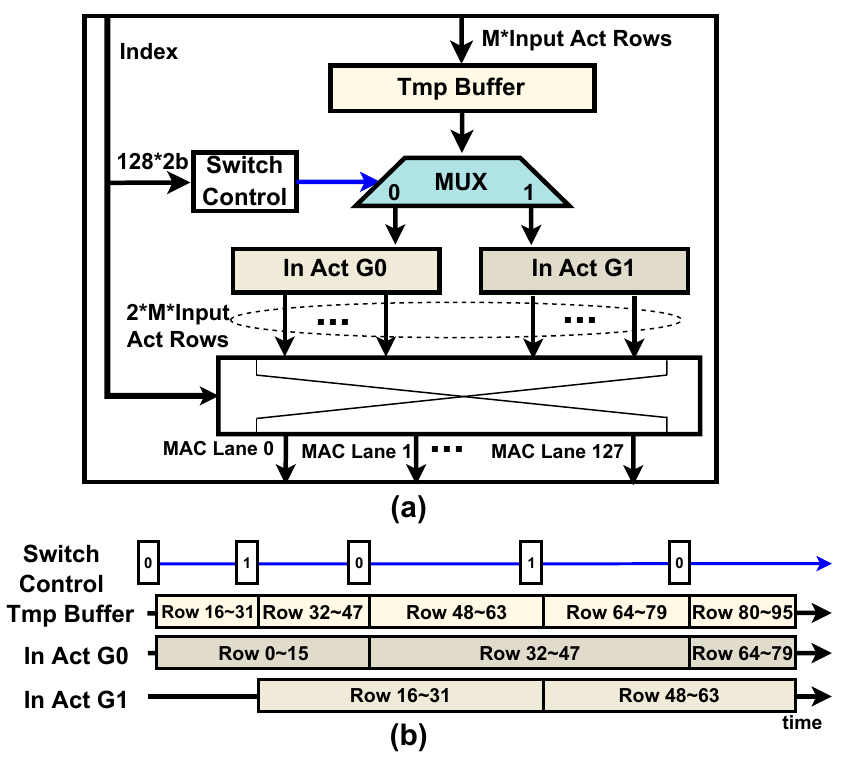}
    \caption{The sequential-write-parallel-read input activation buffer: (a) design scheme and (b) a timing diagram example. }
    \label{fig:input_act_buffer}
\end{figure}

Four classes of reshaping operations are involved in EyeCoD's predict-then-focus processing pipeline: the partition operation (see Fig.~\ref{fig:act_GB} (b)) which tiles one input activation tensor into several partitions along activation feature map dimensions to \yang{enable the sequential processing of the partitions}, the concatenation operation (see Fig.~\ref{fig:act_GB} (c)) \yang{which concatenates several tiled output activation tensors generated sequentially by the MAC lanes to the output tensor by along the channel dimension}, the downsampling operation (see Fig.~\ref{fig:act_GB} (d)) for downsampling layers which drops the activations in each activation feature map, and the upsampling operation (see Fig.~\ref{fig:act_GB} (e)) for upsampling layers which inserts zeros or duplicates activations in each activation feature map.
We propose an activation GB storage arrangement where each activation memory bank address stores one tile of activations along the channel dimension, e.g., 16 activation pixels along the channel dimension per address, 
This activation GB storage arrangement considers the reshaping operation granularities along activation feature map and channel dimensions to simplify controlling designs. 
\hr{We provide a more detailed explanation below. In particular,} this storage arrangement considers that the reshaping operations along feature map dimensions (i.e., the partition, downsampling, and upsampling operations) are usually \hr{at the granularity of 1.} 
In contrast, the granularity of the reshaping pattern along channel dimension (i.e., the concatenation pattern) is related to the number of MAC lanes assigned to a certain layer, which is a multiplication of 16 in our design. 
Fig.~\ref{fig:act_GB} (a) gives an example of the storage arrangement of one activation tensor with the shape of 6$\times$6$\times$24. We place four memory banks in parallel for one activation GB and the 6$\times$6$\times$24 activation tensor takes 24 addresses in total. By properly accessing the activation tiles' addresses, the aforementioned four activation reshaping operations are easily supported.

\textbf{Sequential-write-parallel-read Input Activation Buffer Design.}
The sequential-write-parallel-read input activation buffer design is demonstrated in Fig.~\ref{fig:input_act_buffer} (a) with a timing diagram example in Fig.~\ref{fig:input_act_buffer} (b).
In the sequential-write-parallel-read input activation buffer, a temp buffer sequentially fetches $M$ input activation rows from the Act GBs ($M=16$ in this design) for next round of processing and then stores the fetched rows in two interleaved buffers (i.e., In Act G0/G1) following the design principle in Sec.~\ref{sec:accelerator_challenges_oppotrunities}. After the MAC lanes finish the current round of processing, they can read the input activation rows in parallel from In Act G0/G1. 
Thanks to this sequential-write-parallel-read buffer design, $2\times$ higher bandwidth (i.e., $2\times M$) is achieved without memory access stalls which can satisfy the bandwidth requirement for EyeCoD's predict-then-focus processing pipeline..

\section{Experiments}

In this section, we present a thorough evaluation of the proposed EyeCoD framework, including the experiment setups in Sec. \ref{sec:exps_setup}, the overall benchmark with CPUs/GPUs and previous SOTA eye tracking processors in Sec. \ref{sec:overall_comp}, and the evaluation and ablation studies of EyeCoD's algorithm and accelerator in Sec. \ref{sec:alg_exps} and Sec. \ref{sec:hardware_exps}, respectively.

\subsection{Experiment Setups}
\label{sec:exps_setup}

\textbf{Model, Datasets, and Training Settings.} 
\uline{Model}: We use the RITNet~\cite{PredictiveNet} and FBNet-C\zy{100}~\cite{wu2019fbnet} as our backbone model for eye segmentation and gaze estimation stage, respectively. 
\uline{Datasets:}
For evaluating our proposed predict-then-focus pipeline, we use OpenEDS2019 and OpenEDS2020 dataset~\cite{garbin2019openeds,palmero2021openeds2020} for segmentation and gaze estimation, respectively. OpenEDS2019 segmentation dataset~\cite{garbin2019openeds} consists of around 8916 labeled images for training 2403 images for validation. OpenEDS2020 gaze estimation dataset~\cite{palmero2021openeds2020} consists of around 128,000 labeled images for training and 70,400 for validation. 
\cheng{To simulate the FlatCam reconstructed image, we follow the proposed simulation and reconstruction method in \cite{FlatCam}.} 
\uline{Training Settings:} \cheng{For evaluating the performance of the entire pipeline, we first crop 512$\times$512 image patches from the center of all images in both datasets for satisfying the requirement that FlatCam's input images are square \cite{FlatCam}.}
(1) For eye segmentation, we downsample the input image from \cheng{$512\times512$} resolution to $128\times128$ resolution before feeding it into the model. Following the award-winning solution~\cite{chaudhary2019ritnet}, we train the model for 300 epochs with a hybrid loss consisting of standard cross entropy loss, generalized dice loss, boundary aware loss, and surface loss. We optimize the model using Adam optimizer~\cite{kingma2014adam} with learning rate $1\times 10^{-3}$, and batch size 8. 
(2) For gaze estimation, we resize the input image to $256\times 256$ resolution \cheng{and crop a 96$\times$160 ROI region} before passing it into the gaze estimation network. Following~\cite{palmero2021openeds2020}, we train the model with arccosine loss for 25 epochs using Adam optimizer~\cite{kingma2014adam} with learning rate of $5\times 10^{-4}$, and batch size of 32.

\begin{figure}[t]
    \centering
    \includegraphics[width=1.0\linewidth]{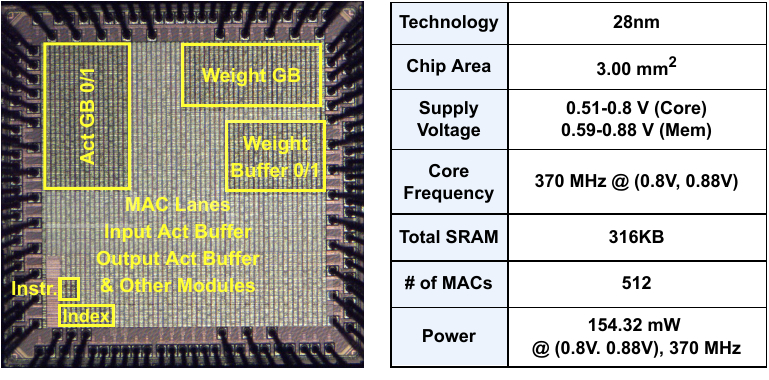}
    \caption{{\name} silicon prototype: die photo and chip specifications.}
    \label{fig:die_specs}
\end{figure}

\begin{table}[t]
\setlength{\tabcolsep}{0.25em}
    \centering
    \caption{Accelerator configurations.}
    \resizebox{\linewidth}{!}{
\begin{tabular}{@{}ccccc@{}}
\toprule
\textbf{Act GB0/GB1} & \textbf{Weight Buffer0/1} & \textbf{Weight GB} & \textbf{Index SRAM} & \textbf{Instr. SRAM} \\ \midrule
512KB * 2 & 64KB * 2 & 512KB & 20KB & 4KB \\ \midrule \midrule
\textbf{MAC Lanes} & \textbf{MACs/MAC Lane} & \textbf{Area} & \textbf{Clock frequency} & \textbf{Power} \\ \midrule
128 & 8 & \multicolumn{1}{c}{8 $mm^2$} & 370MHz & \multicolumn{1}{c}{335mW} \\ \bottomrule
\end{tabular} 
    }
    \label{tab:hw_config}
\end{table}

\textbf{Baselines and Evaluation Metrics.}
\uline{Baselines}: We choose four general computing platforms \hr{EdgeCPU (Raspberry Pi), CPU (AMD EPYC 7742), EdgeGPU (Nvidia Jetson TX2), GPU (Nvidia 2080Ti)}, and one eye tracking processor CIS-GEP \cite{ASIC_gaze_processor} as baselines. The batch size for CPU and GPU is set to 1 for a fair comparison. \cheng{For analyzing the overall estimation accuracy of our FlatCam-based system and the proposed predict-then-focus pipeline, we compare EyeCoD with the winner method in OpenEDS2020 \cite{palmero2021openeds2020}.} 
\uline{Metrics}: We evaluate all above platforms in terms of both throughput and energy efficiency. In addition, we compare the achieved mIOU and FLOPs comparisons for EyeCoD's segmentation models, gaze estimation error in degrees and FLOPs for EyeCoD's gaze estimation models.


\textbf{Hardware Platform Setup.  }
\uline{Silicon-validated EyeCoD.} 
Fig.~\ref{fig:die_specs} illustrates the specifications of the silicon-validated EyeCoD's accelerator which is denoted as the chip for convenience. Specifically, the chip is fabricated in a commercial 28nm HPC CMOS technology, with a total chip area of 3mm$^2$, a core/memory supply voltage of 0.8V/0.88V, and a power of 154.32mW at a 370MHz frequency. The chip is equipped with 316KB SRAM and 512 MACs.
\uline{Evaluation Methodology.} 
\yang{To enable a fair comparison with the baseline designs with a larger area than the silicon-validated chip, we develop an in-house cycle-accurate simulator of {\name}'s accelerator, for which the MAC and memory access costs are derived from the real chip measurement or the post-layout simulation~\cite{zhao2022flatcam}. 
The simulator is verified against the Register-Transfer-Level (RTL) implementation of {\name}'s accelerator to ensure its correctness,}
\underline{Technology-dependent Parameters.} 
Tab.~\ref{tab:hw_config} presents the characteristics of our cycle-accurate simulator of {\name}'s accelerator used throughout the experiments. Specifically, we implemented 128 MAC lanes with each containing 8 MACs. The SRAM includes two Act GBs with 512KB each (Act GB0/GB1 in Fig.~\ref{fig:arch}), two weight buffers with 64KB each (Weight Buffer0/1 in Fig.~\ref{fig:arch}), one weight GB with 512KB (Weight GB in Fig.~\ref{fig:arch}), one index buffer with 20KB (Index SRAM in Fig.~\ref{fig:arch}), and one instruction buffer with 4KB (Instr. SRAM Fig.~\ref{fig:arch}). Same as the silicon-validated chip, the cycle-accurate simulator assumes a 370 MHz frequency.

\subsection{Overall Performance Comparison}
\label{sec:overall_comp}

\begin{figure}[t]
    \centering
    \includegraphics[width=\linewidth]{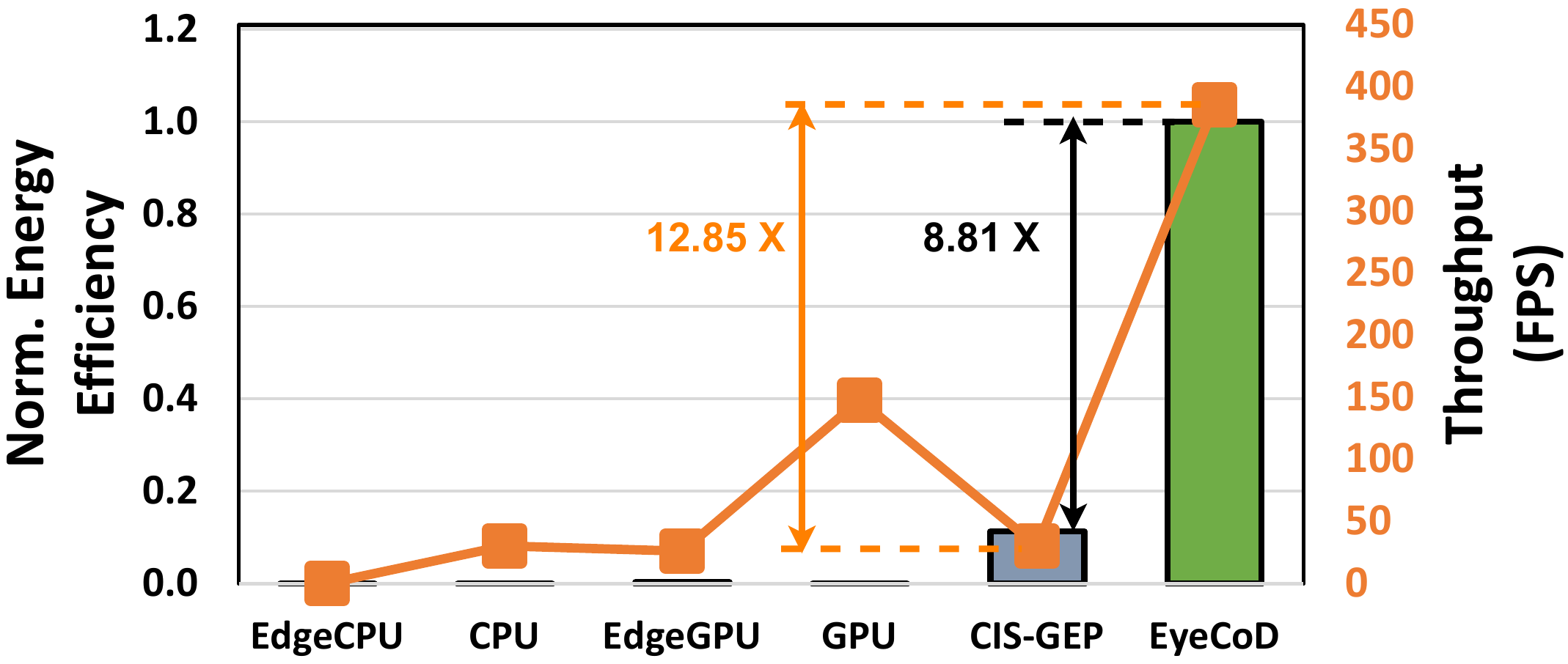}
    \caption{Overall comparison between EyeCoD and other five baselines in terms of both energy efficiency and throughput.
    }
    \label{fig:overall_comp}
\end{figure}

In this part of experiment, we benchmark the proposed EyeCoD's overall performance (normalized energy efficiency vs. throughput) against the  CPUs/GPUs  and previous  SOTA  eye  tracking  processors~\cite{ASIC_gaze_processor}. As shown in Fig.~\ref{fig:overall_comp}, the proposed EyeCoD consistently achieves both the best normalized energy efficiency and throughput among all the baselines. 
\hr{
Specifically, EyeCoD achieves 2966.65$\times$, 12.75$\times$, 14.83$\times$, 2.61$\times$, and 12.86$\times$ improvements in terms of throughput as compared to EdgeCPU, CPU, EdgeGPU, GPU, and CIS-GEP baselines, respectively. 
Meanwhile, EyeCoD maintains a high energy efficiency, achieving 8.81$\times$ improvement as compared to the most competitive baseline ASIC accelerator CIS-GEP \cite{ASIC_gaze_processor}.
}
We conjure that the improved throughput comes from the sensor-processor co-design and the various optimization techniques as proposed in Sec.~\ref{sec:accelerator_architecture}, e.g, customization for different layer types, the proposed sequential-write-parallel-read input activation buffer design, that offer better matched spatial tiling of each operator type to the MAC lanes and reduce the memory stalls, respectively, thus in turn leading to the improved hardware utilization. 
As for the improved energy efficiency, the tailored handling of the depth-wise convolution layer produces much more intra layer data reuse, resulting in much reduced off-chip memory access and thus also reduced power consumption.

\subsection{Evaluation of the EyeCoD Algorithm}
\label{sec:alg_exps}

\textbf{Algorithm pipeline evaluation.}
We first evaluate the necessity of our proposed pipeline with optimized input resolution by benchmarking EyeCoD with vanilla gaze estimation. 
\cheng{For the baseline method, we report the winner in \cite{palmero2021openeds2020} that achieves 2.31 degrees error at the cost of 1.82GFLOPs.
As shown in Tab.~\ref{tab:gaze}, our proposed EyeCoD with the same model (i.e., ResNet18) achieves comparable gaze estimation error (0.10 degree higher) with over 69.2\% FLOPs reduction, suggesting that (1) a FlatCam-based eye tracking system does not degrade the accuracy, (2) it is necessary to use an optimized input size for a higher accuracy-efficiency trade-off. }



\begin{table}[t]
    \centering
    \caption{Benchmark EyeCoD \textbf{gaze estimation} algorithm on OpenEDS'20 dataset with FlatCam reconstructed dataset. The adopted setting in EyeCoD is marked in \textbf{bold}.}
    \setlength{\tabcolsep}{0.20em}
    \resizebox{\linewidth}{!}{
    \begin{tabular}{c|c|c|c|cc}
    \toprule[2pt]
        Model & Camera & Resolution & Error & Parameter & FLOPs \\
        \hline
        ResNet18 & Lens & 224$\times$224 & 3.17 & 11.18M & 1.82 G \\
        \hline
        ResNet18 & \multirow{4}{*}{ FlatCam } & \multirow{4}{*}{ 96$\times$160 } & 3.27 & 11.18M & 0.56G \\
        MobileNet &  &  & 3.43 & 2.23M & 0.10G \\
        FBNet-C100 &  & & 3.23 & 3.59M & 0.12G \\
        \textbf{FBNet-C100 (8-bit)} & & & \textbf{3.23} & \textbf{3.59M} & \textbf{0.01G} \\
        \bottomrule[2pt]
    \end{tabular}
    }
    \label{tab:gaze}
\end{table}

\textbf{Gaze Estimation}
\cheng{On top of the FlatCam system and the pioneering SOTA model ResNet18, we evaluate the effectiveness of various models for gaze estimation. As shown in Tab.~\ref{tab:gaze}, EyeCoD with FBNet-C100 (8-int) (highlighted in bold) improves the error of ResNet18 by 0.04 while reducing FLOPs by 78.2\%.}

\textbf{ROI Prediction.} 
We further validate the effectiveness of the ROI prediction algorithm we propose by benchmarking our eye segmentation algorithm performance under various settings. Compare to the original result in \cite{garbin2019openeds}, our proposed algorithm face three more challenges, (1) lower signal-to-noise ratio (SNR) of the FlatCam reconstructed image, (2) smaller resolution ($\frac{1}{16}$ of the original resolution size), and (3) 8-bit quantized model. 
As shown in the Tab.~\ref{tab:segmentation}, despite the aforementioned challenges and $16\times $ FLOPs reduction, the segmentation algorithm of EyeCoD (marked in \textbf{bold}) still achieves comparable performance (achieving around 93\% mIOU on validation dataset) as the award-winning solution in~\cite{garbin2019openeds}. 
Moreover, when segmenting on the FlatCam dataset instead of the original image, all networks suffers from performance degradation in terms of mIOU ranging between $1.5\%$ to $0.6\%$. However, smaller resolution in general suffers less from the adaption of dataset, we suspect this is due to the relatively lower SNR in FlatCam reconstructed images, making it hard for the models to learn the detailed features in the high-resolution images. 

\begin{table}[]
    \centering
    \caption{Benchmark \textbf{RITNet performance} on OpenEDS'19 dataset under different experiment settings, the adopted setting in EyeCoD is marked in \textbf{bold}.}
    \resizebox{\linewidth}{!}{
    \begin{tabular}{@{}c|c|cc|c@{}}
    \toprule[2pt]
        \multirow{2}{*}{ Model } & \multirow{2}{*}{ Resolution } & \multicolumn{2}{c|}{Eye Segmentation mIOU} & \multirow{2}{*}{ FLOPs } \\
        & & Origin Image & FlatCam Image & \\
        \hline
         U-net & 512$\times$512 & 93.3 & 92.5 & 14.1G \\
        RITNet &          512$\times$512 & 95.1 & 93.6 & 17.0G \\
        RITNet &          256$\times$256 & 94.7 & 93.8 & 4.1G \\
        RITNet (8-bit) &  256$\times$256 & 94.0 & 92.8 & 0.3G \\
        RITNet &          128$\times$128  & 94.1 & 93.5 & 1.0G \\
        RITNet (8-bit) &  128$\times$128  & 93.3 & \textbf{92.7} & \textbf{0.1G} \\
        \bottomrule[2pt]
    \end{tabular}
    }
    \label{tab:segmentation}
\end{table}

\hr{
\textbf{Ablation Studies.} As shown in Tab. \ref{tab:rebuttal_ROI_ablation_1}, the extracted ROIs augment the gaze estimation by centralizing the eye areas, leading to \cheng{9.41} and \cheng{8.24} error reductions as compared to random crop or central crop, respectively. This set of experiments validate that extracting ROIs not only reduces the computational cost but also helps to mitigate the undesired effect of FlaCam's blurred images.
We also test different ROI sampling frequencies, as shown in Tab. \ref{tab:rebuttal_ROI_ablation_2}, and find that (1) the gaze estimation error and the segmentation FLOPs per frame in general gradually increase alone with the increased ROI sampling frequency, while the error reduction is negligible when sampling frequency is higher than 1 over every 50 frames, and (2) a larger ROI size leads to a better gaze estimation accuracy, where the increase is not significant after the ROI size is larger than 96$\times$160. Thus, our adopted setting (i.e., extracting ROI every 50 frames with a size of $96\times160$) achieves an optimal accuracy and inference FLOPs trade-off.  
}

\begin{table}[t]
    \centering
    \caption{\hr{Ablation studies of EyeCoD's ROI prediction.}}
    \resizebox{\linewidth}{!}{
    \begin{tabular}{c|c|c|c}
    \toprule[2pt]
        & Random Crop & Central Crop & ROI (Ours)  \\
        \hline
        Gaze Estimation Error & 12.64 & 11.57 & 3.23\\
    \bottomrule[2pt]
    \end{tabular}
    }
    \label{tab:rebuttal_ROI_ablation_1}
\end{table}

\begin{table}[t]
    \centering
    \caption{\hr{Ablation studies of both EyeCoD's ROI prediction frequency and ROI sizes.}}
    \resizebox{\linewidth}{!}{
    \begin{tabular}{c|c|c|c|c}
    \toprule[2pt]
        ROI Freq. & ROI Size & \tabincell{c}{Gaze Estimation\\Error} & \tabincell{c}{Gaze Estimation\\FLOPs/Frame} & \tabincell{c}{Segmentation\\FLOPs/Frame} \\
        \hline
        25 & 96$\times${160} & 3.23 & 7.58M & 2.5M\\ \hline
        50 & 48$\times${80} & 3.60 & 2.28M & 1.3M\\
        50 & 96$\times${160} & 3.23 & 7.58M & 1.3M\\
        50 & 144$\times${240} & 3.19 & 18.13M & 1.3M\\ \hline
        100 & 96$\times${160} & 3.34 & 7.58M & 0.7M\\
    \bottomrule[2pt]
    \end{tabular}
    }
    \label{tab:rebuttal_ROI_ablation_2}
\end{table}


\subsection{Evaluation of the EyeCoD Accelerator and System}
\label{sec:hardware_exps}



In this experiment, we perform ablation studies to evaluate {\name}'s contributions for better understanding its overall superiority. 
To quantify the impact of different contributions, we build a lens-based system and run the original images of \yang{256$\times$256} resolution on the system. Specifically, the accelerator in the lens-based system removes the hardware-level contributions (including the input activation buffer design, the time-multiplexing workload orchestration, and the intra-channel reuse for depth-wise layers). Please note that the accelerator here keeps {\name}'s input feature-wise partition to fit the same area and adopts the time-multiplexing mode, where one layer of the eye segmentation model and the gaze estimation model occupy the whole MACs iteratively. 
We calculate the impact from each of our contributions by applying the FlatCam sensor and the predict-then-focus pipeline, and hardware-level contributions to the lens-based system one-by-one.

Among the \yang{4.00$\times$} throughput/energy efficiency improvement over the lens-based eye tracking system, adopting the FlatCam sensor and the predict-then-focus pipeline leads to \yang{1.99$\times$} throughput/energy efficiency improvement, while applying the proposed input activation buffer, partial time-multiplexing mode, and intra-channel reuse optimizations further offer \yang{1.22$\times$, 1.28$\times$, and 1.29$\times$} throughput/energy-efficiency improvement, respectively. In particular, 
(1) the proposed predict-then-focus pipeline helps to reduce the image resolution by \yang{76.5\%}, thus improving the throughput as well as the energy efficiency by \yang{1.99$\times$}; 
(2) the sequential-write-parallel-read activation buffer, which enables the parallelism of memory access and processing, helps to reduce input reading stalls due to the limited activation GB bandwidth and thus improves the performance; 
(3) the partial time-multiplexing mode, leveraging the two NNs’ different execution frequencies for workload-orchestration, achieves \yang{1.28$\times$} speedup than the time-multiplexing mode; and 
(4) the intra-channel reuse further reduces \yang{71\%} of the depth-wise layers' processing time, resulting in \yang{1.29$\times$} speedup.    
\begin{table}[t]\centering
\caption{\hr{Throughput and normalized energy efficiency of the proposed EyeCoD w/ and w/o 
predict-then-focus pipeline (P.F.),
sequential-write-parallel-read input activation buffer design (Input.),
partial time-multiplexing workload orchestration (Partial.), 
and intra-channel reuse for depth-wise layers (Depth.).
The last row is the final adopted EyeCoD system.
}}\label{tab:speed_up_break}
\scriptsize
\resizebox{\linewidth}{!}{
    \begin{tabular}{l|c|c}
    \toprule[2pt]
    System$^\mathrm{\star}$  & \tabincell{c}{Throughput\\(FPS)} & \tabincell{c}{Norm.\\Energy Eff.} \\\hline
    Lens-based System$^\mathrm{*}$& \yang{96.34} & 1.00\\
    EyeCoD w/ P.F.$^\mathrm{*}$ & \yang{191.94}  & \yang{1.99}\\
    EyeCoD w/ P.F. \& Input. & \yang{233.64}  & \yang{2.43}\\
    EyeCoD w/ P.F. \& Input. \& Partial.  & \yang{299.04}  & \yang{3.10}\\
    EyeCoD w/ P.F. \& Input. \& Partial. \& Depth.  & \yang{385.66}  & \yang{4.00}\\
    \bottomrule[2pt]
    \end{tabular}
}
\begin{tablenotes}
\footnotesize
\item{$^\mathrm{\star}$ All settings use input feature-wise partition.}
\item{$^\mathrm{*}$ Using time-multiplexing mode.}
\end{tablenotes}
\end{table}





\section{Conclusion}

To this work, we propose, develop, and validate a lensless FlatCam-based eye tracking algorithm and accelerator co-design framework dubbed EyeCoD to enable eye tracking systems with a much reduced form-factor and boosted system efficiency without sacrificing tracking accuracy, targeting next-generation eye tracking solutions. 
On the system level, we advocate the use of lensless FlatCams instead of lens-based cameras to facilitate the small form-factor need in mobile eye tracking systems.
On the algorithm level, EyeCoD integrates a predict-then-focus pipeline that first predicts the region-of-interest ROI
and then only focuses on the ROI parts to estimate gaze directions.
On the hardware level, we further develop a dedicated accelerator that integrates a novel workload orchestration between the aforementioned segmentation and gaze estimation models, and leverages multiple optimization to further improve the acceleration efficiency. 
On-silicon measurement and extensive experiments validate advantages of our EyeCoD in enhancing the end-to-end eye tracking throughput while maintaining the tracking accuracy.

\section*{Acknowledgment}

We would like to acknowledge the funding support from the NSF RTML program (Award number: 1937592) and the NSF EPCN program (Award number: 1934767) for this project.

\bibliographystyle{ACM-Reference-Format}
\bibliography{refs}

\end{document}